\documentclass[11pt]{amsart}
\usepackage[utf8x]{inputenc}
\usepackage[margin=1in]{geometry}
\usepackage{setspace}
\usepackage{color}
\usepackage{graphicx}
\usepackage{epstopdf}
\usepackage{textcomp}
\usepackage{amssymb,amsfonts,amsmath}
\usepackage[math]{iwona}
\usepackage{libertine}

\usepackage[defaultmono,scale=0.8]{droidmono}
\usepackage[T1]{fontenc}
\usepackage{nameref}

\newcommand{\inv}{^{-1}}
\newcommand{\Dist}{\operatorname{Dist}}

\begin{document}
\vspace*{0.35in}

\title[Testing Vascular Scaling Models with MRI]{Testing Foundations of Biological Scaling Theory Using Automated Measurements of Vascular Networks}
\author[MG Newberry, DB Ennis, VM Savage]{
Mitchell G Newberry\textsuperscript{1,\dag},
Daniel B Ennis\textsuperscript{2},
Van M Savage\textsuperscript{1,3,4,*}}

\maketitle

\noindent{\footnotesize 
{\bf 1} Department of Biomathematics, David Geffen School of Medicine, University of California, Los Angeles, Los Angeles, CA, USA
\\
{\bf 2} Department of Radiological Sciences, Biomedical Physics, and Bioengineering, University of California, Los Angeles, Los Angeles, CA, USA
\\
{\bf 3} Department of Ecology and Evolutionary Biology, University of California, Los Angeles, Los Angeles, CA, USA
\\
{\bf 4} Santa Fe Institute, Santa Fe, NM, USA

\medskip

\noindent\dag Current address: University of Pennsylvania, Department of Biology, 433 S University Ave, Philadelphia, PA 19104, USA\\
* vsavage@ucla.edu }
\bigskip

\begin{abstract}
Scientists have long sought to understand how vascular
networks supply blood and oxygen to cells throughout the body.
Recent work focuses on principles that constrain how vessel size
changes through branching generations from the aorta to capillaries and
uses scaling exponents to quantify these changes.  Prominent
scaling theories predict that combinations of these exponents
explain how metabolic, growth, and other biological rates vary
with body size.  Nevertheless, direct measurements of individual
vessel segments have been limited because existing techniques for
measuring vasculature are invasive, time consuming, and
technically difficult. We developed software that extracts the
length, radius, and connectivity of \textit{in vivo} vessels from
contrast-enhanced 3D Magnetic Resonance Angiography.  Using data
from 20 human subjects, we calculated scaling exponents by four
methods---two derived from local properties of branching
junctions and two from whole-network properties.  Although these
methods are often used interchangeably in the literature, we do
not find general agreement between these methods, particularly for
vessel lengths.  Measurements for length of vessels also diverge
from theoretical values, but those for radius show stronger
agreement.  Our results demonstrate that vascular network models
cannot ignore certain complexities of real vascular systems and
indicate the need to discover new principles regarding vessel
lengths.
\end{abstract}


\section{Introduction}
Networks that supply resources are essential to the maintenance and
growth of many natural and engineered systems.  These resource-distribution
networks are pervasive throughout biology. Examples include the tracheal
system in
insects, xylem networks in plants\cite{price2007general}, foraging trails of
ant colonies\cite{latty2011structure,jun2003allometric}, and cardiovascular
systems in
animals\cite{west1997general,murray1926vascular,womersley1955method}. 
Because the cardiovascular system delivers resources and energy to the body, its
structure at least partly determines rates of growth and metabolism
\cite{west1997general,thompson1917growth,savage2008sizing}.  Moreover, the
cardiovascular system plays a role in many diseases --- such as heart disease,
stroke, inflammation, and malignant tumor
growth\cite{jain2005normalization,herman2011quantitative,savage2013using}.
The pervasiveness and importance of material transport in biology motivates the
search for basic principles that help shape resource-distribution networks in
general and the cardiovascular system in particular.

The theory of the structure of vascular networks has roots in the early 20th
century\cite{thompson1917growth,murray1926vascular,womersley1955oscillatory}.
More recent theories predict properties of the entire vascular network by assuming
a hierarchical structure and include those of
West, Brown and Enquist\cite{west1997general}
(henceforth the WBE model),
Banavar et al.\cite{banavar1999size,banavar2014form}, 
Dodds\cite{dodds2010optimal},
Huo and Kassab\cite{huo2012intraspecific}, 
and others\cite{savage2010hydraulic,brown2004toward,price2007general,kolokotrones2010curvature,sperry2012species,allmen2012species}.
These theories assume or predict how vascular structure, such as the radius
of vessels, changes as the network branches from aorta to capillaries. 
While early theories focused on vessel radius, recent theories also incorporate
how vessel length changes across the network\cite{west1997general,dodds2010optimal,huo2012intraspecific,banavar2014form}. 
Many also assume symmetric
branching --- child vessels have identical
properties\cite{huo2012intraspecific,west1997general,sperry2012species,kolokotrones2010curvature}.
In symmetric, hierarchical models, knowledge of both radius and length enables
derivations of how metabolic rate varies with body size across
species\cite{west1997general} and how organismal and tumor growth rate change
with body
size\cite{jain2005normalization,herman2011quantitative,savage2013using}. 

Several theories predict vascular structures will be found to be self-similar
--- some aspect of the network can be viewed as a
rescaled copy of the whole\cite{mandelbrot1977fractals} ---
across specific ranges of 
spatial scale\cite{west1997general,zamir1999fractal}.
Self similarity 
necessarily leads to relationships and distributions that are 
characterized by power laws, whose exponents we call the scaling exponents\cite{schroeder2012fractals}.
Self similarity can either be a strict or statistical property.  
Each new chamber of the nautilus is
a larger exact copy of the previous chamber, while along 
coastlines shorter segments exhibit rescaled statistical properties of the
longer segments\cite{mandelbrot1967long}. Self similarity in nature suggests
scale-free principles
that constrain structure.  An organism may need to maintain
a certain shape at all stages of growth, use a common
developmental program at all scales, or cope with physical
processes with no preferred scale --- such as energy
minimization or turbulence in fluids.  Self similarity 
greatly simplifies many calculations, with applications from
cardiac physiology\cite{zamir2005physics} to allometric
scaling\cite{west1997general}.

Real vasculature is known to deviate from hierarchical, symmetric models in
many ways, leading to criticism and debate about leading
models\cite{dodds2001re,kozlowski2004west}.  Without reliable data, it is
impossible to determine whether these deviations can be ignored, so that
existing theories can accurately predict newly observed
features such as curvature in scaling relationships\cite{kolokotrones2010curvature}.  Price
et al.\cite{price2012testing} recently decried the lack of data for individual
vessel segments that are needed for tests of scaling theory.  Direct
measurements and tests may reject existing principles while laying the
groundwork for the discovery of new patterns and principles in vascular
architecture.  In addition, measurements can parameterize existing equations to
obtain more exact predictions for metabolic rates and growth curves, and can be
used to examine the natural variation in the parameters across species,
tissues, and tumor types.  Vessel segment data preserves the asymmetry,
reticulation, tortuosity, and other features of real vasculature, and
quantitative data about these features may inform future models.  Furthermore,
extreme values or distinct patterns of variation may be signatures of
pathologies that could eventually be used as diagnostics. Recent work involving
direct measurements of plant architecture has begun to realize this
potential\cite{price2011leaf,allmen2012species,mckown2010decoding}.

Comprehensively characterizing vascular structure and obtaining reliable
estimates for vascular scaling exponents requires large numbers of measurements
across orders-of-magnitude in
spatial scale --- ranging from 0.004 mm to 15 mm
for vessel radii in humans. Measurement is complicated because vascular systems
are intertwined with tissues throughout the body at a wide range of spatial
scales\cite{fung1997biomechanics}.
Even gross morphological measurements have
historically taken impressive and time-consuming efforts and
required invasive methods such as
casting\cite{huang1996morphometry, kaimovitz2008diameter,
kassab1993morphometry, zamir1983arterial}.
Zamir\cite{zamir1983arterial} and
Kassab\cite{kassab1993morphometry} constructed explicit
descriptions of small regions of vascular systems --- such as
the coronary artery --- by perfusing fixed specimens with a
silicone or acrylic polymer, dissolving tissue away and
examining each vessel segment under a microscope. These
approaches enabled the measurement of tens of thousands of
vessels from fixed specimens that were then used to test and develop vascular
system models\cite{kassab1993morphometry,zamir1983arterial}.

However, the process of casting may enlarge or damage vessels, and little of
this raw data is publicly available for analysis. In addition, many more
measurements across the range of scales are needed to identify the principles
that shape vascular architecture.  Different physical principles
may dominate at different scales, and mapping out different regimes will
require large amounts of data at each scale.  In the WBE
model\cite{west1997general}, the dominant mechanism of energy loss for blood
flow in the arterioles (radius $\ll$ 1 mm) is viscous dissipation, but near the
heart (vessel radius $\gg$ 1 mm) pulsatile flow and reflection of pressure waves along
vessel walls dominates\cite{zamir2005physics}.  The transitions between these
regimes are neither well-understood theoretically nor described
empirically\cite{savage2008sizing}.  The pulsatile regime --- the focus of our
measurements --- has greater variation in the number of branching orders and
size of vessels across species and is thus the primary determinant of scaling
of metabolic and other vital rates with body size\cite{savage2008sizing}.

Large amounts of vascular data across all relevant spatial scales are
contained within existing angiographic images (e.g., MRI and X-Ray). These
images are obtained non-invasively, thus avoiding problems of damage to
vasculature and allowing for the possibility of longitudinal studies.  The
latent data within these images represents a tremendous opportunity.  All that
is needed is a reliable and automatic method for extracting vascular data from
angiographic images.

We have developed novel software that imports 3D images, creates a
topologically and spatially explicit map of the blood vessel network, and
measures the radius, length, and volume of all visible vessels.  We have
applied this software to 20 magnetic resonance angiograms of living humans to
obtain 3015 data points
 that range in radius from 0.6
- 6.8 mm, representing an order of magnitude. This range corresponds to large
vessels in the pulsatile flow regime relevant to allometric
scaling\cite{savage2008sizing}.  We then analyzed these data based on the four
distinct methods (Eqs~\ref{eq:murray_gen}-\ref{eq:regres}) for measuring
vascular scaling exponents described below.

\section{Model}

Vascular scaling exponents encapsulate how radius and length of vessels change
across the network.  Virtually all scaling relationships for local or global
properties can be expressed in terms of these vascular scaling exponents.
Consequently, we view these scaling exponents as forming the foundation of most
modern biological scaling theory and make them the primary focus of our
analysis.  We here describe four distinct methods for calculating vascular
scaling exponents.

Because the radius of the vessel plays a primary role in determining both
the flow rate and resistance to blood flow through the vessel,
theories for the vascular scaling exponent for vessel radius
often focus on the power to pump blood from the heart to the
capillaries. It is argued\cite{west1997general} that this
power will have been minimized by natural selection to allow
as much power as possible to be available for foraging, growth,
and reproduction. One classical approach minimizes power
loss of blood flow due to viscous dissipation and due to cost of blood
volume in order to derive that flow rate, $\dot{Q}$, depends
on the cube of the vessel radius, $r^3$.  Combining this
result with conservation of fluid flow at a branching
junction yields Murray's law, $r_p^3 = \sum_i r_{c,i}^3$,
where $r_p$ is the radius of the parent vessel segment and
$r_{c,i}$ is the radius of the $i$th child (distal) segment.
Another classical approach for the cardiovascular system is
to minimize wave reflections in pulsatile flow, as Womersley
and West et al.~have
done\cite{womersley1955method,zamir2005physics,west1997general}.
This approach leads to area-preserving branching (or Da
Vinci's Rule), so that the sum of the cross-sectional areas
($\propto r^2$) of child vessels equals the cross-sectional
area of a parent vessel at a branching junction.  In the WBE
model, reflections dominate for large vessels while
dissipation dominates for small vessels ($\ll$ 1 mm).
Moreover, in the WBE model, a volume-servicing
argument\cite{west1997general} is used to derive an
analogous relationship for vessel lengths, $l_p^{3} = \sum_i l_{c,i}^{3}$, 
while Huo and Kassab assume the same relationship but allow the
exponent to vary from length-preserving (exponent of 1) to volume-servicing
(exponent of 3).  

Optimization has been a common approach to developing vascular models
throughout the past century, but it has been highly debated as to which
properties are optimized and what are the tradeoffs among
them\cite{apol2008revisiting,dodds2001re,kozlowski2004west,west1997general,FEC:FEC2022,Savage28122010}.
For instance, Banavar et al.\cite{Banavar07092010} optimize for efficient transport within
three-dimensional bodies and Dodds\cite{dodds2010optimal} minimizes network volume as required to
continually supply metabolites within a body.  Indeed, different principles and
assumptions lead to a variety of relationships between the flow rate and vessel
radius.  Consequently, we express a generalized form of Murray's law 
\begin{subequations}
\begin{equation}
\label{eq:murray_gen} 
r_p^{1/a} = \sum_i r_{c,i}^{1/a}.  
\end{equation} in
which we define the vascular scaling exponent, $a$, for vessel radius to be
consistent with the notation of Price et al.\cite{price2007general}.  The
analogous generalization for vessel length is 
\begin{equation}
\label{eq:volserv_gen} 
l_p^{1/b} = \sum_i l_{c,i}^{1/b} 
\end{equation} 
\end{subequations}
where $b$ is the vascular scaling exponent for vessel length.  

To ease computation, many models further assume that vascular networks are
strictly self-similar and symmetrically branching---child vessels all have
identical properties within a branching level.  In this case, scale factors and
associated scaling exponents can be defined for each branching level, $k$, which
represents the number of branching junctions from the heart to that vessel.
Following the notation of the WBE model, the scale factors are
$\beta=r_{k+1}/r_{k}$ and $\gamma=l_{k+1}/l_{k}$.  For dichotomous branching,
we can solve for $\beta$ and $\gamma$ using Eqs~\ref{eq:murray_gen} and
\ref{eq:volserv_gen}, to find
\begin{equation}
\label{eq:bgqs}
\beta = {r_{k+1}\over r_k} = 2^{-a} \mbox{ and }\gamma = {l_{k+1}\over l_k}  = 2^{-b}.
\end{equation}

Furthermore, for these idealized networks, the frequency distributions of radius
and length follow power laws with scaling exponents $1/a$ and $1/b$. Because
there are $N=2^k$ vessels of radius $r = r_0\beta^k$, Eqs~\ref{eq:bgqs} give
the two power-law relationships 
\begin{equation}
\label{eq:powlaw}
N = (r/r_0)^{-1/a} = (l / l_0)^{-1/b}.
\end{equation}  
Similarly, for any vessel, its radius and length are related to the number of downstream endpoints (e.g. capillaries),
$N_d$, by 
\begin{equation}
\label{eq:regres}
r \propto N_d^{{a}} \mbox{ and } l \propto N_d^{{b}}.
\end{equation}

Eqs~\ref{eq:murray_gen}-\ref{eq:regres} constitute four methods of
calculating vascular scaling exponents. Each method relies on different types
and levels of information.
First, for each branching junction, the generalizations of Murray's law
(Eq~\ref{eq:murray_gen}) and volume servicing (Eq~\ref{eq:volserv_gen})
can be numerically solved for the exponents $a$ and $b$ by Newton's method.
The solution provides a direct local measurement of the scaling exponents at
every junction that we call \textit{conservation-based} scaling exponents. The value is
undefined if a child vessel has radius or length greater than its parent.
Second, for each parent-child pair of vessel segments, the ratio of vessel
radius and length can be calculated. By Eq~\ref{eq:bgqs}, these scale
factors can be used to compute a second local measure --- our \textit{ratio-based}
scaling exponents.
Third, across all vessels and junctions, empirical distributions of radii and
lengths can be fitted to power laws to produce what we term the
\textit{distribution-based} scaling exponents, as in Eq~\ref{eq:powlaw}.
Fourth, across all vessel segments, log-log regressions of radii and lengths
versus the number of downstream endpoints can be performed to derive
\textit{regression-based} scaling exponents, following Eq~\ref{eq:regres}.
These latter two methods each provide single values for the vascular scaling
exponents, $a$ and $b$, across the whole network, and they do not rely on
information about the geometry of individual branching junctions.

In the literature, these four methods for measuring scaling exponents are often
used
interchangeably\cite{bentley2013empirical,price2007general,price2012testing,west2009general}.
However, these are only proven to be identical for symmetrically-branching,
strictly self-similar networks.  Furthermore, it is unknown whether measurements of vascular
scaling exponents using these four methods
(Eqs~\ref{eq:murray_gen}-\ref{eq:regres}) will produce values that are
approximately similar or significantly different.  If they differ, this raises
questions about which of these four measures of vascular scaling exponents, if
any, best corresponds to the scaling relationships predicted by ideal networks
or observed empirically for metabolic and growth rates.

\section{Results}

To begin to answer these questions, we now report results obtained by applying
our new software, angicart, to 20 contrast-enhanced 3D Magnetic Resonance Image
volumes for human head and torso.  We collected 3015
segments across 1473 branching
junctions.  Of the junctions, 1422
were recorded as dichotomous and 51
as trichotomous.  The number of branches on all paths (subjects pooled) between
the aorta and the smallest observable vessels was typically between 3 and 10
(middle 68\%).  Each 3D image volume, corresponding to
151 segments on average, took about
2.7 minutes of CPU time on a single
2.3GHz Intel processor.  As described in the Methods, we used an automated
circularity criterion as an indicator of possible errors in order to omit
vessel identification errors. Our results are based on radius, length, and
volume measurements from only the 1240
segments for which we saw no indication of potential error.

The complete, raw output of our software before any filtering or analysis is
available as \nameref{S1_Dataset}.  The software source code and original
imagery enabling the exact reproduction of this dataset are available online as
a repository in the git revision control system at
\texttt{https://github.com/mnewberry/angicart}.

\subsection{Conservation-Based Exponents}
We attempt to solve Eq~\ref{eq:murray_gen} numerically
at each branching junction.  We label the parent at each branching
junction by traversing the vascular network starting at the
vessel segment of greatest radius, which we assume is a
segment in the aorta.  
If all children are smaller than the parent, the equation is guaranteed to have
exactly one solution, and numerical convergence to the
solution is fast.  Otherwise, the equation may have zero,
one, or two solutions.  Such cases may represent real
anatomical variation, or misidentification of the parent
vessel due to errors or ambiguous topology such as the
Circle of Willis.  We consider only the simplest case,
in which children are smaller than their parent, because
otherwise the solutions are difficult to interpret in the context of existing
vascular network models.  All child radii are smaller than the parent in
82\%
(222) of junctions, and all child lengths are smaller in only
35\%
(94) of junctions.

\begin{figure}[t]
\setkeys{Gin}{width=4.5in}
\includegraphics{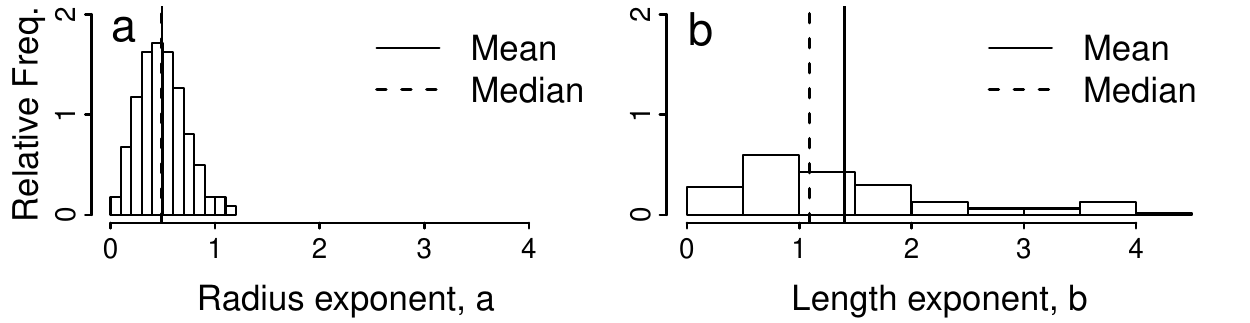}
\caption{{\bf Conservation-Based Exponent Distribution.}
Plots of (a)  the frequency distribution of the vascular scaling
exponent $a$ for vessel radius from solutions to Eq.~(\ref{eq:murray_gen})
using empirical measurements of vessel radii extracted from magnetic resonance
angiography using our software, and (b) the analogous frequency distribution of
vascular scaling exponent $b$ for vessel segment length.}
\label{fig:qdist}
\end{figure}

We show distributions of these measures of the vascular scaling exponents, $a$
and $b$, across branching junctions in Fig~\ref{fig:qdist}. The arithmetic
mean and associated 95\% confidence intervals for these conservation-based
measurements of $a$ and $b$ are presented in Table~\ref{tab:all}.

\begin{table}[!ht]
\begin{tabular}{l c c c c c c}
{\bf Radius Exponent Measures}
  & $N$
  & $a$ 
  & 95\% CI
  & $\sigma$
  & $R^2$\\
\hline
Conservation (Node) &
222 & 
0.49 &
$\pm$ 0.03 &
0.23 &
--\\
Ratio (Node) &
703 & 
0.43 &
$\pm$ 0.03 &
0.39 &
--\\
Distribution (Network) &
657 &
0.30 &
$\pm$ 0.04 &
-- &
0.90\\
Regression (Network) &
1240 &
0.41 &
$\pm$ 0.02 &
-- &
0.66\\
{\bf Theory} & & & & &\\
\hline
WBE Small Vessels & -- & 0.33 & -- & &\\
WBE Large Vessels & -- & 0.50 & -- & &\\
Banavar et al.\cite{banavar2014form} & -- & 0.50 & -- & &\\
Huo and Kassab\cite{huo2012intraspecific} & -- & 0.33--0.50 & -- & &\\
Murray's Law & -- & 0.33 & -- & &\\
\\
{\bf Length Exponent Measures} & $N$ & $b$ & 95\% CI & $\sigma$ & $R^2$\\
\hline
Conservation (Node) &
94 & 
1.40 &
$\pm$ 0.20 &
0.98 &
--\\
Ratio (Node) &
703 & 
0.17 &
$\pm$ 0.12 &
1.57 &
--\\
Distribution (Network) &
518 &
0.73 &
$\pm$ 0.10 &
-- &
0.68\\
Regression (Network) &
1240 &
0.94 &
$\pm$ 0.05 &
-- &
0.19\\
{\bf Theory} & & & & &\\
\hline
WBE All Vessels & -- & 0.33 & -- & &\\
Banavar et al.\cite{banavar2014form} & -- & 0.50 & -- & &\\
Huo and Kassab\cite{huo2012intraspecific} & -- & 0.33--1.00 & -- & &\\
\end{tabular}

\bigskip

\caption{{\bf Measures and theoretical values of the scaling exponents $a$ and
$b$.} $N$ denotes the number of measurements incorporated in the average or
slope.  The 95\% CI indicates the range of confidence on the average or slope.
For node-level measures, $\sigma$ indicates the standard deviation of
measurements made for individual junctions or parent-child pairs for
conservation- and ratio-based measures respectively.  For network-level
(regression- and distribution-based) measures, $R^2$ denotes the correlation in
the fit.  For distribution-based measures, this correlation is between the
natural log of bin size and mean vessel dimension (i.e., radius or length) in
the bin.  For regression-based measures, this correlation is the correlation
between the natural log of dimension and the natural log of the number of
downstream endpoints.  For conservation-based measures, $N$ counts all
junctions of three or more well-segmented vessels (see
\nameref{sec:soft_and_alg}) for which Eqs \ref{eq:murray_gen} or
\ref{eq:volserv_gen} had a solution.  For ratio-based exponents, $N$ counts all
parent-child pairs of vessels in which both the parent and child are
well-segmented. For distribution-based exponents, $N$ counts the number of
well-segmented vessel segments exceeding the minimum-size threshold described
in \nameref{sec:data_fit}.  For regression-based exponents, $N$ counts all
well-segmented vessels.  The 95\% CIs are derived in a manner appropriate to
each method: They are 1.96 times the standard error on the mean for
conservation- and ratio-based measures and the confidence interval on the
SMA regression slope for regression-based measures.  For distribution-based
measures, they are the range of the middle 95\% of slopes derived from
alternative binning as described in \nameref{sec:data_fit}.}
\label{tab:all}
\end{table}

\subsection{Ratio-Based Exponents}
\label{sec:ratexponents}
\begin{figure}[t]
\setkeys{Gin}{width=4.5in}
\includegraphics{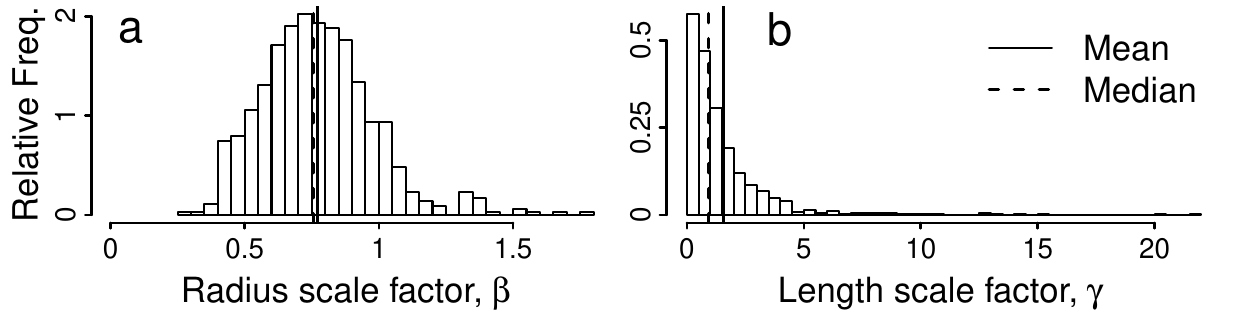}
\caption{{\bf Ratio-Based Exponent Distribution.}
Plots of (a) frequency distribution of scale factor $\beta$, the
ratio of child to parent radii, and (b) the analogous distribution of scale
factor $\gamma$, the ratio of child to parent lengths.}
\label{fig:betadist}
\end{figure}

Topological information allows us to compute $\beta$ and
$\gamma$ directly for each parent-child pair of vessel
segments. Our dataset contains 703 pairs of
parent and child segments with dichotomous branching.  A small proportion of
the $\beta$ and $\gamma$ values may be over-estimated due to
misidentification of the parent-child relationship (see Methods).  This bias
would produce underestimates in $a$ and $b$, but because misidentification is very infrequent, the magnitude of this bias is expected to be within
the measurement error.  The distribution of $\beta$ and $\gamma$ is
displayed in Fig~\ref{fig:betadist}, and the arithmetic mean and associated 95\% confidence intervals of the ratio-based vascular scaling exponents calculated using Eq.~(\ref{eq:bgqs}) are shown in Table~\ref{tab:all}.

\subsection{Distribution-Based Exponents}
\begin{figure}[t]
\setkeys{Gin}{width=4.5in}
\includegraphics{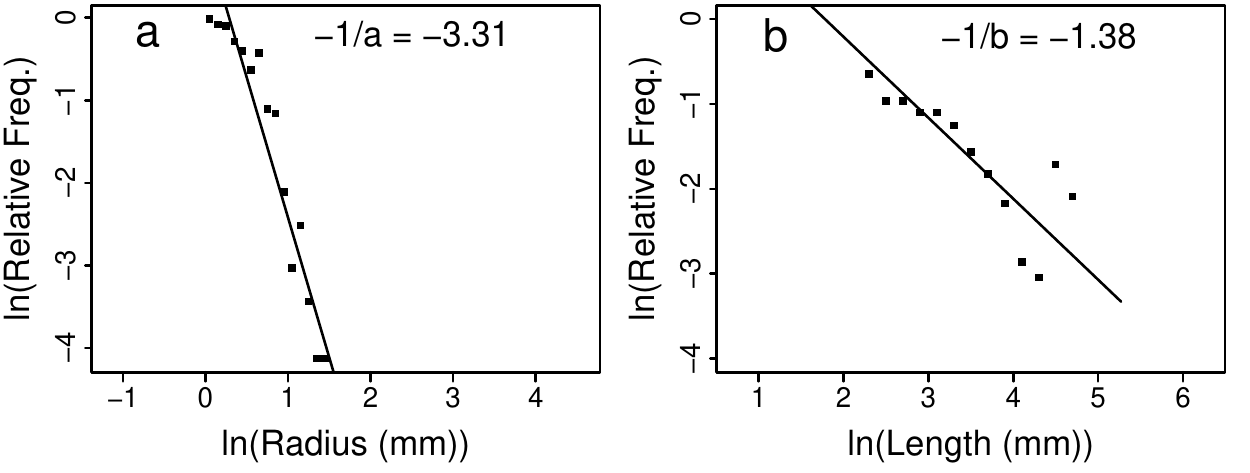}
\caption{{\bf Distribution-Based Exponent Fits.} Standard Major Axis regression
(see Methods) of the natural log of relative frequency (probability density)
against (a) the log of radius, $\ln r$, and (b) the log of length, $\ln l$.  Fit
lines and slope values are shown.  The correlation coefficient $(R^2)$ for the
log relative frequency versus the logs of radius and length respectively are
0.90 and 0.68.}
\label{fig:rdist}
\end{figure}

For symmetrically branching, self-similar networks, the frequency of radius and length
measurements follow power-law distributions.  
We did a linear fit to the log-log transformed histograms of radius and length
measurements (the log of Eq.~(\ref{eq:powlaw})) using SMA
regression. We derived empirical 95\% confidence intervals by
resampling with different bin sizes and cutoff values for the tail (see
\nameref{sec:data_fit}).  The fits are shown in
Fig~\ref{fig:rdist}.  The scaling exponents obtained from our
fits are given in Table~\ref{tab:all}.

\subsection{Regression-Based Exponents}
\label{sec:regexponents} 
\begin{figure}[t]
\setkeys{Gin}{width=4.5in}
\includegraphics{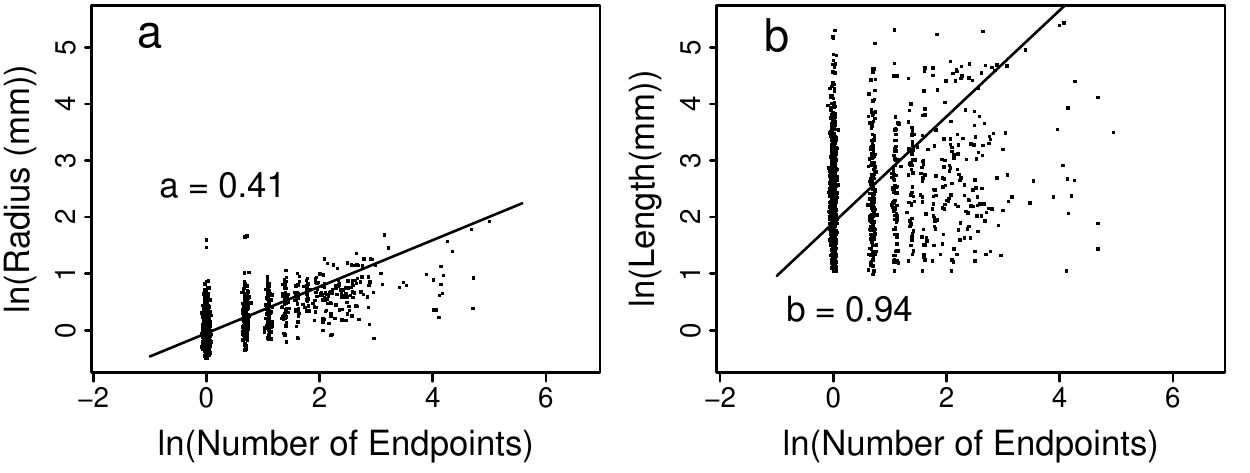}
\caption{{\bf Regression-Based Exponent Fits.}
Standard Major Axis regression (see Methods) of (a) the log of radius, $\ln r$, 
and (b) the log of length, $\ln l$, against the log of the number of downstream
tips, $\ln N_d$.  Fit lines and slope values are shown.  Correlation
coefficients $(R^2)$ are 0.66 and 0.19,
respectively ($P < 0.01$ for each).
}
\label{fig:betanreg}
\end{figure}

By taking the logarithm of Eq.~(\ref{eq:regres}), we can estimate
the vascular scaling exponents, $a$ and $b$, by performing
regressions of the logarithm of the number of downstream tips
$\ln N_d$ against the logarithm of radius, $\ln r$, and the
logarithm of length, $\ln l$, respectively
\cite{bentley2013empirical}.  Regression lines are shown in
Fig~\ref{fig:betanreg}.  The measured slopes, which are the
estimates of the vascular scaling exponents, and associated 95\% CI are shown in Table~\ref{tab:all}.

\subsection{Sensitivity Analysis}

As in all analyses based on experimental data, measurement errors affect
uncertainties in quantities calculated from the raw data.  Consequently, we
investigate the sensitivity of our calculated scaling exponents and entire
analysis to the choice of threshold intensity in our algorithm as well as to
Gaussian noise in the image quality.

\begin{figure}[t]
\setkeys{Gin}{width=4.5in}
\includegraphics{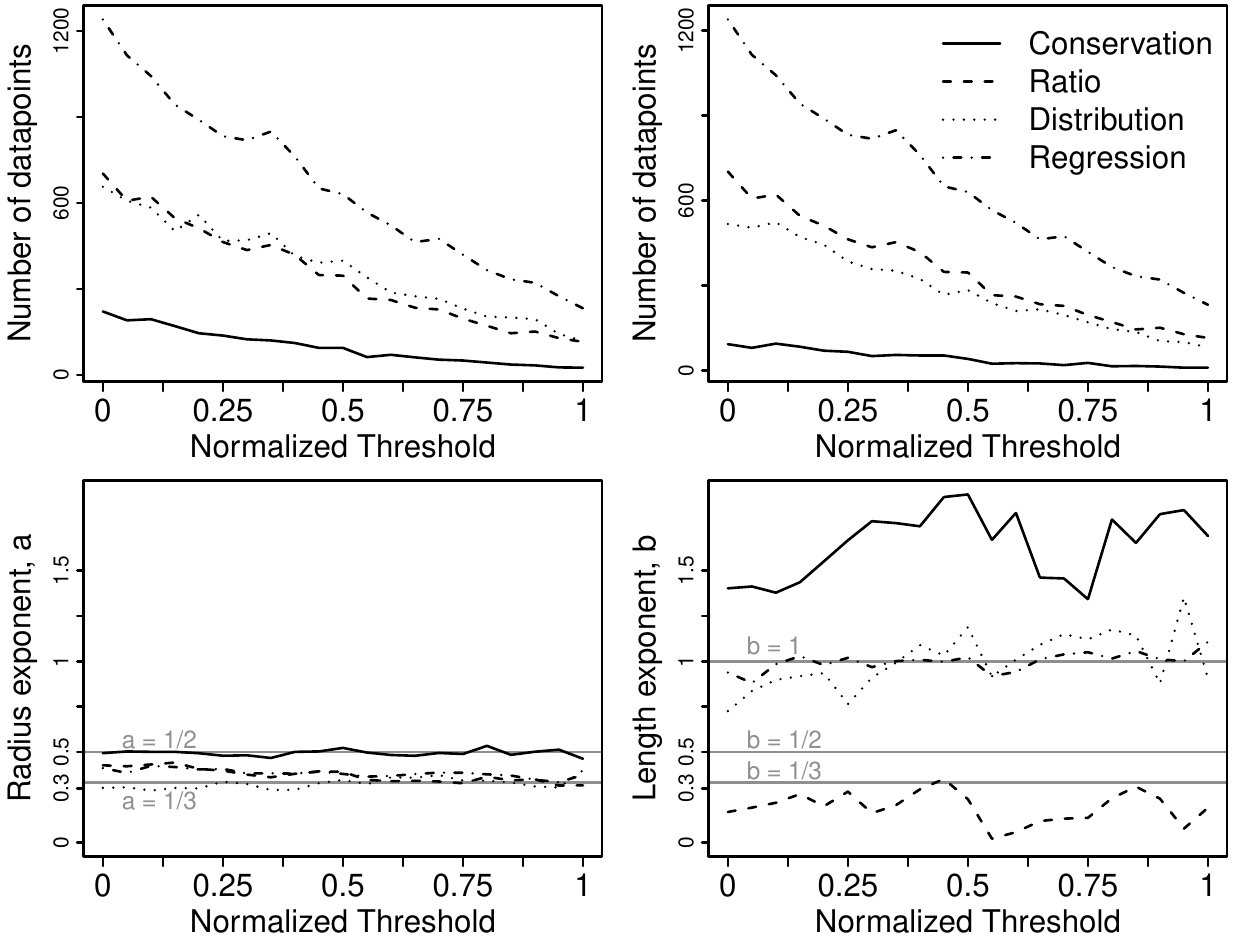}
\caption{{\bf Sensitivity of the measured scaling exponents to the threshold
intensity values that select which voxels are part of the vessel lumen.}  We
normalized the threshold so that the minimum threshold (0.0) is the value used
in our results and figures above, while the maximum threshold (1.0) is the
largest value for which at least 30 vessel segments are visible. We re-scaled
the threshold values because the raw value is different for each image. The top
left panel shows the number of vessel segments used to calculate the four
scaling exponents for vessel radius versus the normalized threshold value, as
in the $N$ column of Table~\ref{tab:all}. The top right panel shows the number
of vessel segments used to calculate the four scaling exponents for vessel
length versus the normalized threshold value, as in the $N$ column of
Table~\ref{tab:all}. For both of these top two panels, the number of data
points decreases with threshold value as it must. The bottom left panel depicts
how the four calculated scaling exponents for vessel radius vary with
normalized threshold value.  The bottom right panel depicts how the four
calculated scaling exponents for vessel length vary with normalized threshold
value.  Horizontal lines indicate points of comparison to theory as described
in Table~\ref{tab:all}.}
\label{fig:thresh_sens}
\end{figure}

An intensity threshold is used in our algorithm to select the voxels from the
image that describe the shape of the vessel lumen.  Lower thresholds reveal
more vessels, so for our analysis above, we used the minimum threshold that
produced reliable segments, as described in \nameref{sec:soft_and_alg}
and \nameref{S1_Text}.  Because the threshold affects the boundaries of the vessel
lumen, the vessel radius also depends on the threshold. For each image there
is a minimum acceptable threshold that we used above, and for our sensitivity
analysis, we also chose a maximum threshold to be the largest value for which
at least 30 vessel segments are visible.  For our plots in Fig~\ref{fig:thresh_sens}, we normalized the threshold to range from 0 and 1.
That is, 0 is the threshold used for the results above, and 1 is the maximum
threshold for which at least 30 vessel segments are visible.  For each
normalized threshold increment of 0.05 between 0.00 and 1.00, we ran our
entire analysis and calculated the four scaling exponents for radius and
length (Fig~\ref{fig:thresh_sens}). The values of the scaling exponents at
a normalized threshold of 0.00 recapitulate Table~\ref{tab:all}.  Our results
remain qualitatively similar as we increase the threshold and include fewer
segments.  However, at higher normalized thresholds, we can no longer resolve
differences between some exponents that are resolvable at normalized threshold
0.00, as expected from the law of large numbers.

Noise in images may also cause errors in the identification of vessels or in
estimates of the radius or length of vessels.  We conducted a sensitivity
analysis similar to the above by adding Gaussian noise that varied in magnitude
from 0.47 (the measured baseline level of noise in the foreground of one image)
to 4.7\% (10 times the baseline noise level) of the maximum voxel intensity.
Results (\nameref{fig:noise_sens}) show no significant changes in our
results with higher levels of noise.

As another measure of uncertainty, we located vessels with radius estimates
that differed with threshold despite high reliability in vessel identification
(vessel endpoints were similar across at least 5 threshold values). We located
12 vessel segments from 9 different patients that matched this criterion.
Across these 12 vessels, the mean radius estimate varied from 0.9 to
6.7mm, but the coefficient of variation (=standard deviation/mean across
thresholds) ranged only from 0.02 to 0.08. Moreover, the coefficient of
variation was uncorrelated with radius (Pearson correlation = 0.07), suggesting
that individual vessel radius measurements are precise to roughly $\pm10\%$ (2
times the coefficient of variation) regardless of vessel size.  For a specific
threshold, we calculate how the measured vessel radius differs from the mean of
the vessel radius across all thresholds. At each threshold, this difference
tended to be in the same direction (mostly positive or mostly negative) across
the 12 vessels we measured. Moreover, the magnitude of the difference was
proportional to vessel radius (most $R^2 > 0.99$). Consequently, the ratio of
vessel radii at any specific threshold is roughly equal to the ratio of the
means of the vessel radii across thresholds and also equal to the ratio of
vessel radii at any other specific threshold. That is, within the plausible
range of thresholds that we explore here, the calculated scaling ratios are
largely independent of the choice of threshold value, implying that the choice
of threshold does not create any bias in the calculated scaling exponents or
ratios.

Of course, our sensitivity analysis cannot exclude all possible sources of
error or bias.  In the Methods sections, we discuss how our results might be
affected by other possible sources of error, such as tree topology
identification, the patient population, small vessel censoring, vessel lumen
misidentification, skeleton line selection, centerline quantization, and
vessel segment misattribution. 

The positional accuracy of MRI is high, so errors arise due to classification
or interpretation of voxels.  Because the threshold parameter and noise in the
image primarily control how we classify voxels --- the first step of analysis
--- these are the major determinants of subsequent errors. In our analysis, as
presented in Figs~\ref{fig:thresh_sens} and \nameref{fig:noise_sens}, no
systematic biases are observable, so we conclude that our results are highly
robust to the largest and most notable sources of uncertainty.

\subsection{Comparing Measurements with Each Other and
Predictions} The estimated values of vascular scaling exponents
obtained using our four different methods are all presented in
Table~\ref{tab:all}.  All pairs of measures are statistically
significantly different (Welch's t-test, $P < 0.01$) except for
the ratio-based and regression-based $a$ ($P =
0.18$).  Values of $a$ based on
conservation rules at branching junctions, scale factors for
parent-child pairs, and regression of $\ln r$ versus $\ln N$ are
all between $a = 1/2$ (the WBE prediction for large vessels) and
$a = 1/3$ (Murray's law, the Banavar et al.
prediction\cite{banavar2014form}, and the WBE prediction for
small vessels), and the remaining distribution-based $a$ includes
$a=1/3$ in its 95\% CI. The conservation-based exponent $a$ is
not statistically-significantly different from $a = 1/2$.

Different measures for $b$ range from $0.17$ to $1.40$ and all are
statistically significantly different ($P < 0.01$) from each other and from the
volume-servicing and area-servicing values of $b = 1/3$ and $1/2$ respectively.
It is notable, however, that the difference between regression-based and
distribution-based measurements of $b$ is no longer resolvable at normalized
thresholds higher than 0. The distribution-based and regression-based exponents
lie between area-servicing and length-preserving.  These discrepancies between
measures suggest that vessel segment lengths are poorly modeled by strictly
self-similar and symmetrically branching networks.

\section{Discussion}

Our software acquired direct measurements of a large number of connected vessel
segments from \textit{in vivo} angiography.  We calculated vascular scaling
exponents in these data using four methods to directly compare values from real
vascular networks with each other and with theoretical values from the WBE
model, Murray's Law, Banavar et al.\cite{banavar2014form}, and Huo and
Kassab\cite{huo2012intraspecific}.  Intriguingly, our results lead to
contrasting conclusions for the changes in vessel radius and length across
scale. 

For the vascular scaling exponent $a$ that quantifies changes in the radius,
the conservation-based and ratio-based estimates are closer to $a=1/2$ than
$a=1/3$. These estimates support work on large vessels by West et al., Zamir
and Banavar\cite{west1997general,zamir2000vascular,banavar2014form}, in
contrast to Murray's law, which does not distinguish large and small vessels.
West et al.  derive that the dominant source of power loss for large vessels
(estimated to be $r \gg$ 1 mm) is the reflection of pressure waves at
branching junctions, while for small vessels, power loss is dominated by
viscous dissipation between blood and the vessel walls.  Because of the
resolution of our MRI volumes, we are able to extract data mostly for vessels
with a radius greater than 1 mm, corresponding to the large vessel regime in
the WBE model. Consequently, our results for the scaling of radii are
supportive of the area-preserving branching of large vessels, corroborate
recent findings for plants\cite{bentley2013empirical}, and reject the
possibility that Murray's law might apply, either generally or on average, to
junctions of large vessels.  This result demonstrates that minimizing
energy-dissipation due to blood flow does not capture the guiding principles
that shape the vascular system across all scales. Future studies using higher
resolution angiography (e.g., micro-CT) to obtain data for small vessels are
needed to test Murray's law and the WBE prediction for small vessels and to
determine if minimizing energy dissipation is a relevant principle at any
scale.

For the vascular scaling exponent, $b$, for vessel lengths, the discrepancies
between predicted and estimated values are more difficult to reconcile and
interpret.  None of the four measures of $b$ agree with each other or provide
support for volume- or area-servicing branching, while only the
regression-based method provides support for length preservation.  Within the
WBE model, $b$ is predicted to be $1/3$ based on an argument that the vascular
network must be volume-servicing for the entire body\cite{west1997general}.
This volume-servicing argument has been questioned on theoretical
grounds\cite{kozlowski2004west}, and here we provide empirical evidence that
volume-servicing or any other conservation law for length does not hold locally
at branching junctions.  Indeed, for the conservation-based exponents, 65\% of
branching junctions violate the model so severely that exponent values are
undefined.  The volume-servicing argument is supposed to apply across at least
the vast majority of scales and is a key element of the WBE explanation for the
$3/4$ allometric scaling relationship between metabolic rate and body size.
The breakdown between this argument and the real vascular networks we measured
may occur because, contrary to the WBE argument, the length of a vessel segment
is not a reliable indicator of the volume it services.  The correlations
between length and number of downstream endpoints, a proxy for volume serviced,
is very low compared to the same correlation for radius (0.2 versus 0.7).
Considering only the largest vessels, this is not surprising.  The ascending
aorta is only a few centimeters in length and services most of the body, while
the carotid artery is much longer (at least 10 cm in length) and services
only half the head.  Our results imply that either modification of the
volume-servicing argument is needed or some new principle yet to be discovered
guides the distribution of vessel lengths as the vascular network branches
throughout the body. These new developments could lead to corrections to the
power-law predictions of the original WBE theory that may agree better with
recent findings of ``curvature'' in the allometric
relationship\cite{kolokotrones2010curvature}.

Beyond the differences discussed thus far, vessel lengths and radii also differ
in their distributions for vascular scaling exponents and scale factors.
Measurements of $a$ and $\beta$ exhibit a strong central tendency
(Figs~\ref{fig:qdist} and \ref{fig:betadist}), while the scale factor for
length, $\gamma$, has a highly skewed distribution with typical values that are
not well-described by the mean.  Thus, a derivation implicitly based on a
mean-value approximation may be successful for predicting vessel radii but fail
to predict scaling relationships involving vessel lengths.  Thus, while
hierarchical symmetric models may fail outright to adequately describe vessel
lengths, the discrepancies between vessel radii in real networks and idealized
models, such as the WBE model, may only result in minor corrections to model
predictions.  This may help explain the success of the WBE model in predicting
a wide range of phenomena.

Differences in results for vessel radius and length could be tied to different
strengths of the constraints on vessel geometry.  Radii and length
distributions have previously been observed to differ in the external branching
of plants and leaves\cite{price2012scaling,bentley2013empirical}. One
explanation for this is that viscous power loss depends much more sensitively
on vessel radius (as a $4$th power, $\propto r^4$) than on vessel length
(linearly, $\propto l$). Thus, the strength of selection for optimal vessel
radius is much stronger than for optimal vessel length, implying evolution has
more often sacrificed vessel length when negotiating tradeoffs in anatomy.
Another potential explanation is that vessel radii are self-similar due to a
local constraint at each branching junction, whereas vessel lengths may be
constrained only at larger scales -- organs and organisms -- that more
accurately capture how the vascular network needs to span and feed a spatially
inhomogeneous body.

Disagreement about the value of the length scaling exponent between our four
methods indicates that assumptions of the simplest model must be violated so
strongly as not to hold even approximately.  That is, strict self similarity,
symmetric branching or both must be strongly violated for the real vascular
networks we measured.   Our data reveal pervasive asymmetry in both radius and
length between child vessels. How far the results for symmetric networks
generalize to asymmetric networks has been explored very
little\cite{turcotte1998networks}.  The differences we observe between
different measures of vascular scaling exponents could be explained by the
inability of existing theories to account for the asymmetry of real vascular
networks.  Developing a theory to account for asymmetric branching may be
challenging. For instance, accounting for asymmetry would require at least two
scale factors for radii (e.g., $\beta_{big}$ and $\beta_{small}$) and two for
lengths (e.g., $\gamma_{big}$ and $\gamma_{small}$).  These additional scale
factors and associated scaling exponents would necessarily change our analysis
and our estimates for the ratio-based scaling exponent, and would potentially
change our interpretation of the distributions of $\beta$ and $\gamma$.  Rather
than thinking of distributions of $\beta$ and $\gamma$, we would think of joint
distributions of $\beta_{big}$ and $\beta_{small}$, for example.  For similar
reasons, our interpretation and analysis of the frequency distribution of
radius and length could be altered, thus affecting the estimates of the
distribution-based exponents as well.  Angicart outputs $\beta$ and $\gamma$
values for each vessel pair, and can provide the detailed information required
for future studies of multiple scale factors for length or radius and
asymmetric branching.

All of the models discussed in this paper ignore any reticulation or loops in
the vessel topology, in contrast to recent work on leaf venation
networks\cite{mileyko2012hierarchical,blonder2011venation,katifori2010damage,corson2010fluctuations}.
Our analysis also follows this assumption.  However, loops are known to occur
anatomically in healthy (Circle of Willis) and diseased (tumors, arteriovenous
malformations) tissue.  Extensions to our software and to theory could address
this issue.  Such an extension could be used to investigate abnormal tumor
vasculature\cite{jain2005normalization}, or allow new theory to be developed to
explain the normal anatomical function of reticulation.  There are also other
spatial aspects that have received theoretical attention, such as branching
angle, that our software is already capable of recording.  Many more tests
could be performed with data on microvasculature.  For instance, Huo and
Kassab\cite{huo2012intraspecific} have published scaling relationships for how
crown volume and length change with stem radius.  Testing these requires
knowledge of the full crown, down to the microvasculature.  Similarly,
Dodds\cite{dodds2010optimal} makes predictions for virtual vessels that
coincide with real vessels only at the smallest scales.

We developed new software and applied it to MRI of human head and torso to
obtain one of the most detailed datasets for examining branching architecture
in vascular networks. In addition, we conducted a comprehensive data analysis
that uses both local and global methods to measure scaling exponents. Together,
this new software, data, and analysis provides valuable information for
answering fundamental questions about vascular system morphology. The public
release of our imaging software, angicart, should enable researchers to ground
future vascular network theories in empirical data. The software facilitates
comparison across spatial scales and between studies by operating uniformly on
all tomographic imaging methods. Because imaging can be done non-invasively,
our method affords the opportunity to record all spatial information of
\textit{in vivo} vasculature through time or across development. 

We explain four different methods to estimate vascular scaling exponents
from spatially-explicit data. Although researchers use and sometimes
interchange these four methods, we found that all four methods can lead to
different results, and that for scaling exponents for vessel lengths, these
differences can be dramatic. This result is in stark contrast to theoretical
calculations for idealized, symmetric networks that predict all methods will
give identical values. We advise caution when interpreting different methods
and estimates as the same scaling exponent because this could lead to
misperceptions and disagreements among studies. For instance, regression-based
estimates are the most common across levels of biological organization while
distribution-based estimates are used for forests\cite{west2009general}, so
comparing these estimates to each other must be done with care. The differences
we observe call for a new understanding of the relationships between the local
geometry of vessels and the global properties of vascular systems. New theory
should be developed to accommodate the anatomical variation and asymmetric
branching we observe in real networks. 

\section{Methods and Materials}

\subsection{Ethics Statement}
After local institutional review board approval and written
informed consent had been obtained, 20 consecutive adult
patients with clinically suspected supraaortic arterial
occlusive disease were prospectively enrolled to
evaluate new MRI methods for the study of carotid
atherosclerosis\cite{lohan2007mr}.

\subsection{Image Acquisition}
\label{sec:image_acq}
Although other data was recorded in that study, we use only the images.  In
that study less than 0.5\% of observed vessel segments had notable
luminal narrowing, so we conclude patient selection and
enrollment did not affect our results.
We acquired contrast-enhanced magnetic resonance angiograms
(CEMRA) of the upper torso, neck, and head in the 20 human
subjects (N=20) using a 3 Tesla Siemens Trio scanner
(Siemens Medical Solutions, Erlangen, Germany). The data
acquisition details have been previously
described\cite{lohan2007mr}. In brief, the CEMRA images were
acquired after an antecubittal vein injection of gadolinium
based contrast agent (Gd-DTPA, Magnevist, Bayer Shering
Pharma AG, Berlin, Germany).
The image volumes have dimensions that are typically close to
$380 \times 640 \times 128$ voxels, with each voxel nearly
isotropic and between $700 \times 700 \times 800$ \textmu{}m and
$800 \times 800 \times 900$ \textmu{}m.  The resolution and imaged
volume are typical of high-quality 3T MRI.  The point-spread-function for MRI is known to be precise and equivalent to the programmed pixel size\cite{liang2000principles}.  In practice, the geometric accuracy is known to be sub-millimeter\cite{scheib2004high}.  The vessel networks
in each image are clearly visible due to the sharp image
contrast provided by the presence of the contrast agent,
which makes the blood appear bright relative to dark non-blood
tissues. We averaged each $2 \times 2 \times 2$-voxel
cube of adjacent voxels into a single 1.4-1.8 mm voxel to remove noise, reduce processing time, and match conventionally-acquired resolutions.  This reduced noise-induced errors
without substantially changing the number of vessel segments
represented.
In two of our 20 image volumes, segmentation failed because
bright, non-blood tissues were present very close to the blood
volume, so that no threshold value excluded all non-blood objects
as described in \nameref{S1_Text}.  We did not record any
vessel segments from these image volumes.  We saw no relationship
between failure of segmentation and vascular system geometry.

\subsection{Software and Algorithm}
\label{sec:soft_and_alg}
\begin{figure}[t]
\setkeys{Gin}{width=4.5in}
\includegraphics{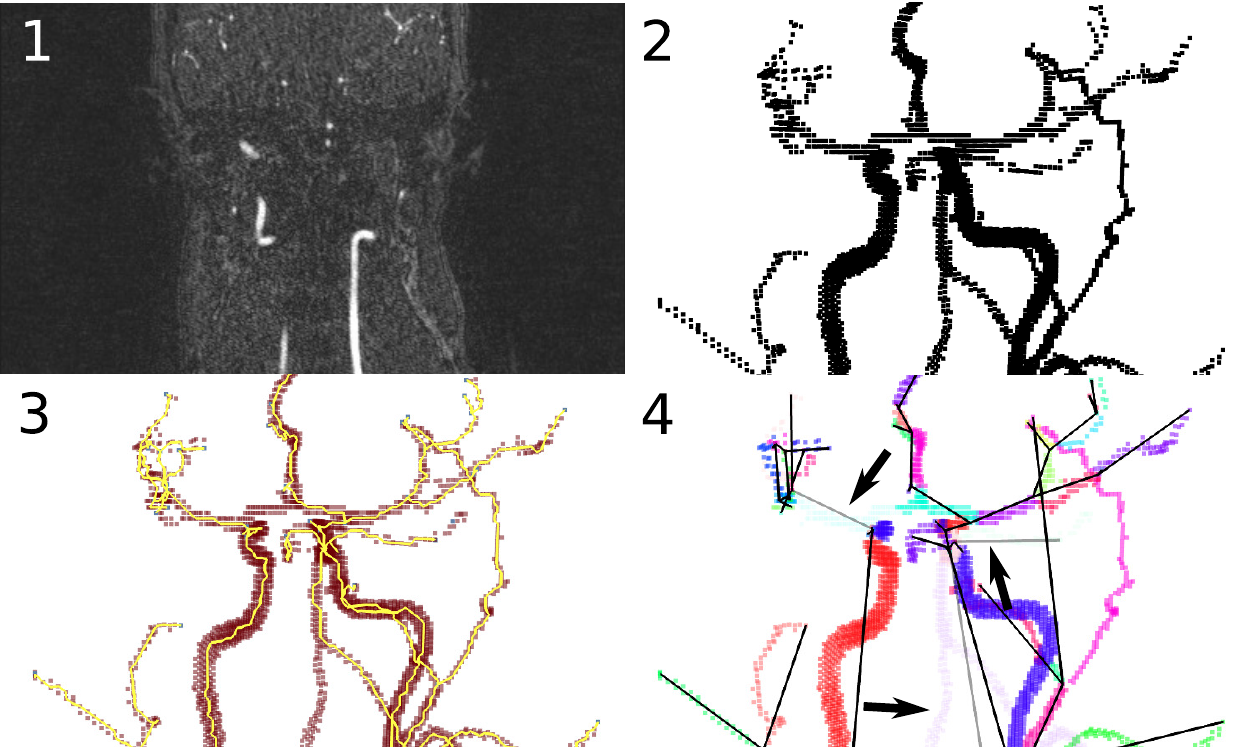}
\caption{{\bf Images from each step of the automatic vessel segmentation
process implemented by our software, angicart.} 1.  Midplane 2D slice from MRI
input (original imagery).  The input to the software is any tomographic
imagery, regardless of imaging method.  MRI, CT, and MicroCT are all possible
data sources.  2. The network mask (3D rendering of included points). This is
the largest group of connected pixels that exceed an intensity threshold. 3.
The skeleton.  Angicart skeletonizes the vessel network by removing any (red)
points that do not disconnect the endpoints (pale blue).  The points that
remain after this process is completed are the centerline (yellow).  Branch
points are those with more than two neighboring voxels, and endpoints are those
with only one neighboring voxel.  4. Vessel segment decomposition.  Angicart
partitions the skeleton into vessel segments. Each segment is colored randomly,
and black lines are used as a simple map of the endpoints of each segment.
Some segments (depicted translucent and indicated by black arrows) are ignored
by the analysis, in this case due to reticulation present in normal human
anatomy.}
\label{fig:overview}
\end{figure}

We created a free, open source software package --- angicart ---
to read tomographic images of vascular networks, to automatically
decompose a vessel lumen into vessel segments, and to measure the
geometry and topology of the segments
(Fig~\ref{fig:overview}).  The software and data used in this
study are available on the
internet (\texttt{https://github.com/mnewberry/angicart/}) under a
GNU Public License.  Our software starts by classifying voxels in
a 3D image as part of the vessel lumen if they are within the
largest connected group of voxels that exceed an intensity
threshold\cite{tsai2009correlations,price2011leaf}, as with other
level set or thresholding methods.  We use a manual binary search to select the
threshold value that best matched visual identification of vessels.  The result
is a 3D binary image, called the \textit{network mask}.  Next, we use spatial
criteria to find the endpoints of the vessel network, where the
vessels become too small to detect.  Given these endpoints, we
find the centerline and branch points of vessels by
skeletonization\cite{kirbas2004review,zhou1999efficient} (See
\nameref{S1_Text}).  We implement skeletonization using an
erosion technique -- successively removing voxels until no more
are removable without disconnecting the endpoints in the
mask\cite{fisher2005dictionary}.  The voxels that remain after
erosion lie within approximately 1 voxel-width of the true
centerline.  

This information allows us to partition the network mask into
segments by attributing each voxel to the segment whose
centerline is closest to it.   We record how vessels are
connected and measure the length of each segment's centerline and
the volume of each segment.  Following a geometric argument (see
\nameref{S1_Text}), quantization error leads us to
overestimate length. We compute radius as $r = \sqrt{V/\pi
l}$.  Radii are therefore underestimated on average due to the
quantization error in length measurements. This bias affects the
accuracy of individual length and radius measurements, but does
not bias estimates of scaling exponents (relative measures) as
long as the percent error does not change systematically with
vessel size, which it does not.  As a final filter of possibly
misclassified vessels, we omit vessels in which more than 20\% of
the voxels lie further than $(r + 1)$ from the centerline, or
whose total volume is less than 4 voxels.  Further details of
each step are presented in the \nameref{S1_Text}.

\subsection{Data Fitting}
\label{sec:data_fit}
We determined the conservation-based node scaling exponents by solving
Eq.~(\ref{eq:murray_gen}) numerically using Newton's method implemented in
OCaml\cite{leroy2012ocaml} and iterated until the sum of powers was within
0.00001 of 1.  We estimated the regression-based scaling exponents using
Standard Major Axis (SMA) regression of the natural log of radius (length)
against the natural log of the number of downstream
endpoints\cite{warton2006bivariate}. We used SMA regression because the
variability and uncertainty in the y-axis (vessel radius or length) is as large
as the variability and uncertainty in the x-axis (number of downstream
endpoints).  Furthermore, SMA is appropriate because our goal is to obtain the
best estimate of the scaling exponents (slopes) and not the best prediction of
$y$ given $x$.

We estimated distribution-based scaling exponents by fitting the
tail of the probability distribution of radius and length to a
power law.  That is, we binned log-transformed data and
determined the slope of the log of probability density versus the
log of radius and length using SMA regression\cite{warton2006bivariate}.  We used 20 bins
and discarded 5 and 7 initial bins of radius and length
respectively.  Blood vessels near the resolution limit of MRI may
not be visible.  Although dimensions measured from observed small
vessels are used in our other methods, counts of small vessels
are unreliable due to censoring.  Thus, we discarded initial bins
in order to exclude vessel sizes where non-uniform censoring of
values might occur.  We computed standard errors by varying bin
size and the number of initial bins discarded (up to $\pm 3$
each) and using the middle 95th percentile of these values.  By
binning our data and fitting the power law using SMA regression, we
avoid problems that can arise when using maximum-likelihood estimators to fit
our power-law distributions. 
Specifically, the maximum-likelihood estimators are derived with specific
assumptions and support choices such as smooth and continuous or discrete and
integer, as in Clauset et al.\cite{clauset2009power}.  In contrast, our
distributions may be somewhere in between: continuous with an increased
likelihood to take values near certain points, such as powers of $\beta$ times
the aorta radius.  Our SMA regression on bins of simulated vessel data produced
stable estimates with relatively little bias in comparison to fits based on
published maximum-likelihood estimators.

\section{Acknowledgments}

We thank J. Paul Finn and Derek Lohan for providing the MRI image data used in
this study.  We thank Elizabeth Sweedyk, Robert Zinkov and Thomas Bushnell,
BSG for helpful discussions of the algorithms, datastructures and programming
of angicart. We thank Joshua Plotkin and David McCandlish for helpful conversations.  We are grateful to Cormac McCarthy for comments on the
manuscript.


\pagebreak

\section*{Supporting Information}

\subsection*{S1 Text}
\label{S1_Text}
{\bf Detailed Methods and Sensitivity Analysis.}

\subsection*{Network Mask}
\label{sec:seg}
The first step in our process is to obtain a binary mask (or
set of voxels) that describes the shape of the blood vessel
network.  Separating a structure of interest from the
background is known in the computer vision literature as
segmentation\cite{fisher2005dictionary}.  The literature on image segmentation
is voluminous and includes studies specifically examining
segmentation of tomographic images of blood vessels.  There
are many approaches to segmentation of blood vessels,
including region-growing, ridge-tracing, watershed models,
spectral approaches, and deformable
contours\cite{kirbas2004review}.  We chose a threshold-based
approach for its conceptual simplicity, estimability of
errors, and lack of implicit assumptions about blood vessel
geometry.  There are more sophisticated approaches, but they
rely on circularity\cite{aylward2002initialization},
volume-servicing\cite{jiang2010vascular}, or other
assumptions to improve their visual quality.  
We avoid these assumptions as 
they might bias the output in favor of models our
measurements are designed to test.  By using simple
thresholds rather than Bayesian classification, we remain
agnostic to the image capture method by ignoring the physics
of the imaging system.

To select a mask representing the blood vessel network, we
calculated the set of voxels that passed an image-specific
intensity threshold, then chose the largest connected group of
such voxels.  To identify the largest connected group, and at
many other steps of the analysis, we consider a set of voxels as
an adjacency graph.  In an adjacency graph, the nodes are the
voxels, and the edges are connections between neighboring nodes.
We considered two voxels to be connected if they were adjacent in
any sense, that is, the voxels touch at at least one corner.  The
distance along the edge is the distance between the center points
of neighboring voxels.  In the language of graph theory, the
largest connected group of voxels that pass a given threshold is
the largest connected component of the adjacency graph.  A
typical network mask is shown in panel 2 of
Fig.~(\ref{fig:overview}).  We use only the largest connected
component because smaller groups of voxels are either noise,
patient motion artifacts, fatty tissue, or isolated vessels whose
relationship to the rest of the network cannot be determined.
Finding the largest connected component is the rate-limiting
process at this stage, but its time is still $O(N\ln(N))$ for an
$N$-voxel image.

The threshold parameter affects how much of the network is
visible. The value affects the volume measurement of each vessel
segment to some degree, but the magnitude is expected to be
small\cite{tsai2009correlations}. At lower (more permissive)
thresholds, more of the network is visible, but dimmer objects
that are not blood are more likely to be part of the largest
connected component.  Fortunately, these misidentified objects
are readily identifiable to a human observer, allowing us to
easily choose a threshold at which no such objects appear.  The
optimal threshold is the lowest threshold that does not
misidentify any objects.  We chose an optimal threshold for each
image using a manual binary search.    

\subsection*{Endpoint Identification}
\label{sec:tips}
The skeletonization process removes voxels from the network
mask until only those required for maintaining a single
connected component remain.  Without some initial set of
non-removable nodes, this process would remove
all nodes.  Therefore, it is important to reliably determine
a set of non-removable nodes -- network endpoints that
represent the most distal visible part of each vessel
branch.  Given the network endpoints, skeletonization will
reduce the network mask to centerlines, but skeletonization
itself cannot identify the endpoints, because they may be
removed without disconnecting the graph.  Failure to
identify endpoints before skeletonization results in the
loss of vessel segments from detection (see Methods).

We identify endpoints as the local maxima of a distance
transform starting from an interior voxel.  Distance
transforms are discussed in the next section.  This transform
assigns higher values to voxels that are more distal in the
network, measured along the contours of vessels.  Voxels at
a local maximum of distance are the network endpoints.  We
consider a local maximum any voxel whose distance value is
greater than all of its neighbors.  Since this sometimes
leads to multiple endpoints in a single terminal vessel, we
collapse endpoints that are within the largest vessel radius
(7 mm) of one another and joined by a straight line
through the network mask.

\subsection*{Skeletonization}
\label{sec:skel}
We use skeletonization to identify vessel centerlines and
branch points.  A skeleton is an irreducible set of voxels
that connect a set of endpoints.  That is, the
removal of any non-endpoint voxel in the skeleton breaks its adjacency graph
into at least two connected components.  
Thus, we compute our skeleton by recursively
removing (eroding) voxels from the network mask, provided
that they do not break the new mask into two components.
There typically
does not exist a unique skeleton for a given network mask
and set of endpoints.  We chose a particular skeleton in
which the remaining voxels conform to the blood vessel
centerlines by preferentially removing nodes from the outside (surface) of the
mask first.

To remove outermost voxels first, we use a distance
transform to rank the voxels according to a measure of how
close they are to the outside of the mask.  We chose a
measure that is lower for voxels closer to the exterior of
the mask, and lower when a voxel has more neighboring voxels
outside the mask.  By removing voxels with more outside neighbors first, we remove voxels at convex points on the surface, smoothing the surface during erosion.
A distance transform $D(v)$ of a
voxel $v$ has the property that 
\begin{equation}
\label{eq:dt}
D(v) = \min_{n \in N(v)} \{\Dist(v,n) + D(n)\},
\end{equation} 
where $N(v)$ is the set of $v$'s 26
neighbors and $\Dist(v,n)$ is the Euclidean distance from
$v$ to a neighbor $n$, which depends on whether the neighbors share one, two, or four corners (ie, a face).  This property is valid except when $v$ is the origin, or part of an initial boundary condition.  Given a value $D(v)$ at any set of voxels, this
property (\ref{eq:dt}) can be used to
extrapolate the distance transform to any connected set of
voxels using a greedy algorithm.  We use an analog of
Dijkstra's algorithm\cite{sniedovich2006dijkstra}.  We supply the values $D(v)$ at points on the surface --- the boundary condition.
We choose the boundary to be the set $B = \{ v$ such that $N(v)
- M \ne \emptyset\}$ of voxels with any neighbors outside
the network mask $M$, and define $D(v)$ on the boundary to be
\begin{equation*}
\left(\sum_{n \in N(v) - M} {1 \over \Dist(v,n)}\right)\inv.
\end{equation*}  This
inverse sum of inverses assigns lower scores to voxels with
more neighbors outside the network mask.

Erosion requires checking each voxel for whether its removal
disconnects the remaining voxels.  Calculating the number of
connected components of the remaining voxels is computationally
expensive ($O(N\ln N)$), making the erosion process
notoriously slow\cite{zhou1999efficient}.  However, we
greatly speed our algorithm by exploiting the fact that
removal of a voxel cannot disconnect the whole graph unless
it disconnects its neighbors.  Checking the last condition
is fast ($O(1)$).  A few passes of removing such voxels
eliminate most of the voxels in the network, creating a
close approximation to the skeleton.  To erode the last few
removable voxels, we use the whole-network algorithm,
calculating connected components for each remaining voxel.
This novel algorithm produces a skeleton in a small fraction
of the time.

Because the skeleton is not unique, different erosion
algorithms result in different skeletons.  For our 20
images, our erosion leads to adequate representations of
vessel centerlines.  We expect any erosion algorithm that
preferentially removes voxels roughly according to the above
criteria to well-represent the centerlines and give
similar results. 

\subsection*{Segment decomposition}
\label{sec:seg_decomp}
Using the skeleton, we can detect which voxels correspond to the
branch points of the blood vessel network.  The skeleton itself
is a set of voxels, and hence its adjacency graph is not a tree
because small cycles occur near branch points.  Thus, we first
compute a minimal spanning tree of the skeleton.  In this tree,
nodes with more than two neighbors exactly correspond to branch
points, and nodes with only one neighbor are vessel endpoints.
Branch points and endpoints demarcate vessel segments.  We
consider the length of a vessel segment to be the distance along
the tree between the segment's ends.  This distance is the sum of
distances between the centers of adjacent voxels, which are
either $1$, $\sqrt{2}$ or $\sqrt{3}$ times the voxel width.  This
path length along the voxel grid is greater than or equal to the
Euclidean distance between two points, and thus quantization of
the vessel centerline will cause lengths to be overestimated in
comparison to other studies which do not require vessel contours
to conform to the grid\cite{antiga2004robust,kline2010accuracy}.
Much as with the Manhattan metric (discrete $l_1$ norm), the
ratio between the grid distance and Euclidean distance is
independent of the scale of the grid.  That is, at any given
orientation relative to the grid, the bias is a fixed ratio
independent of line length.  This error is correctable with
smoothing, but because our use of length measurements is to
determine scaling exponents (relative measures), our results are
not affected by biases that increase length of vessels by a
constant factor.  Consequently, we choose to ignore biases of
this type.

Given the centerlines of vessel segments, we can attribute each
voxel in the network mask to the vessel segment whose centerline
lies closest to it.  To accomplish this, we use Dijkstra's
algorithm to generate a shortest path tree in which any path
along the center lines is zero distance.  Removing all branch
points then breaks the shortest path tree into one connected
component per vessel segment.  The connected component of each
segment contains all information about the segment size, shape,
and spatial position.  The volume of a segment is the number of
voxels multiplied by the volume of each voxel. We compute the
radius of each segment from the length and volume as $r =
\sqrt{V/\pi l}$.  This is effectively equivalent to averaging
multiple measurements of radius along the vessel, resulting in a
low error in radius.  We also label vessel segments and record
their topology to compute the number of downstream endpoints,
scaling exponents $a$ and $b$, and scaling ratios $\beta$ and $\gamma$.

Segment decomposition leads to erroneous vessel segments
when: 1. a closely-spaced bundle of vessels is identified as
a single vessel; 2. endpoint identification misses an
endpoint; 3. the network mask contains a loop; or 4. patient
motion artifacts cause large volumes of blood to 
appear as patches which are skeletonized as individual blood
vessel segments.  Case 1 introduces segments with
erroneously large radii into the distributions, but it
occurs rarely, and usually in conjunction with a loop.
Cases 2 and 3 cause one segment to be missed, and its voxels
attributed to adjacent segments, but these malformed
segments are detectable.  Most of the voxels of tubular
vessel segments are within a distance from the centerline
less than the average vessel radius.  We eliminate any
vessel in which more than 20\% of the voxels are further
than $(r + 1)$ from the centerline, or whose total volume is
less than 4 voxels.  This criterion also eliminates most of
the vessels erroneously introduced in case 4.

\pagebreak

\subsection*{S1 Fig}
\label{fig:noise_sens}
{{\bf Sensitivity of the measured scaling exponents to added image noise
ranging from 1 to 10 times the baseline rate (0.47\%).}  The top left panel
shows the number of vessel segments used to calculate the four scaling
exponents for vessel radius versus the magnitude of the added noise. The top
right panel shows the number of vessel segments used to calculate the four
scaling exponents for vessel length versus the magnitude of added noise.  The
bottom left panel depicts how the four calculated scaling exponents for vessel
radius vary with the magnitude of added noise. The horizontal lines indicate
the WBE predictions for large and small vessels.  The bottom right panel
depicts how the four calculated scaling exponents for vessel length vary with
the magnitude of image noise. The horizontal line indicates the WBE prediction.}

\vspace{\baselineskip}

\setkeys{Gin}{width=4.5in}
\includegraphics{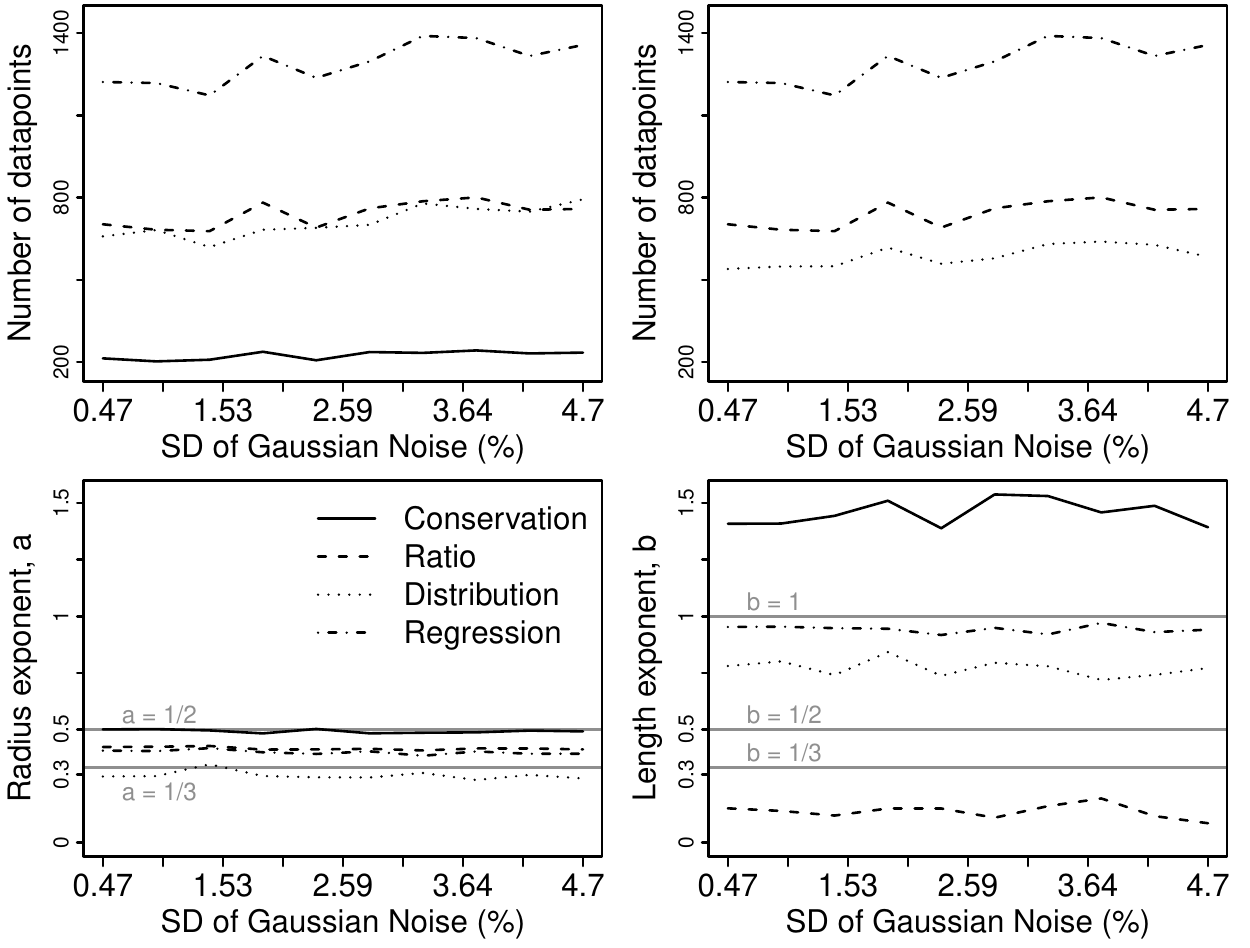}

\vspace{2in}

\subsection*{S1 Dataset}
\label{S1_Dataset}
{\bf Vessel Segment Dataset.} The complete, raw output of angicart on the 18
images used in the study, in tab-separated values (tsv) format is available at\\
\texttt{https://github.com/mnewberry/angicart/raw/example/example.dicom\_small.all.tsv}

\end{document}